\documentstyle[12pt]{article}

\def\be{\begin{equation}}
\def\ee{\end{equation}}
\def\ba{\begin{array}{c}}
\def\ea{\end{array}}
\def\p{\partial}
\def\ben{$$}
\def\een{$$}
\begin{document}

\titlepage

\begin{center}{\Large \bf
Complex Calogero model
 with real energies
 }\end{center}

\vspace{5mm}

\begin{center}

Miloslav Znojil\footnote{ e-mail: znojil@ujf.cas.cz} and Milo\v{s}
Tater\footnote{ e-mail: tater@ujf.cas.cz}

 \vspace{3mm}

\'{U}stav jadern\'e fyziky AV \v{C}R, 250 68 \v{R}e\v{z}, Czech
Republic\\

\end{center}

\vspace{5mm}

\section*{Abstract}

Recently, ${\cal PT}$ symmetry of many single-particle
non-Hermitian Hamiltonians has been conjectured sufficient for
keeping their spectrum real.  We show that and how the similar
concept of a ``weakened Hermiticity" can be extended to some
exactly solvable two- and three-particle models.

\vspace{9mm}

\noindent PACS 03.65.Ge, 03.65.Fd

\vspace{9mm}

\begin{center}
 {\small \today, ptcal.tex file}
\end{center}

\newpage

\section{Introduction}

Insight into bound states in quantum mechanics is facilitated by
solvable models.  They clarify the structure of the
single-particle wave functions in the context of supersymmetry
\cite{Khare}, Lie algebras \cite{Ushveridze} and Sturm-Liouville
oscillation theorems \cite{Pruefer}. This approach can immediately
be extended to the systems of more particles where a very
exceptional role is played by the Calogero's exactly solvable
Hamiltonian \cite{Calogero}
 \ben
 H^{(A)} = - \sum_{i=1}^{A}\
\frac{\p^2}{\p{x_i}^2}
 + \sum_{ i<j=2}^{A}\left [
\frac{1}{8}\,\omega^2   \,(x_i-x_j)^2
 +\frac{g}{
 (x_i-x_j)^{2}}\right ]\ .
  \een
It describes $A$ particles on a line with $2m = \hbar =1$ and its
Lie algebraic treatment proves enormously productive
\cite{Turbiner}. The analysis and interpretation of some of its
properties is also facilitated in the non-singular limit~$g \to 0$
where the interaction degenerates to the mere harmonic-oscillator
attraction.

Recently, the so called ${\cal PT}$ symmetric quantum mechanics of
Bender et al \cite{BBjmp} offered a new picture of some
single-particle models. For example, the most common spectrum of
the harmonic oscillator in $D$ dimensions  was re-interpreted as a
special case of a non-equidistant real spectrum pertaining to a
slightly more general non-Hermitian model~\cite{ptho}.

The picture is based on a complexification of coordinates which
breaks the Hermiticity of the Hamiltonian but does not destroy the
reality of the energies. In the present paper we intend to show
that such a complexification method can be generalized and applied
to some many-particle Hamiltonians. In a constructive way we are
going to demonstrate the existence of non-Hermitian modifications
of $H^{(2)}$ and $H^{(3)}$ which consequently preserve the reality
of the spectrum.


\section{Acceptable solutions}

At any integer $A\geq 1$ the introduction of the centre of mass
 \ben
R=R^{(A)} = \frac{1}{\sqrt{A}}\ \sum_{k=1}^{A}\,x_k
 \een
enables us to eliminate the bulk motion of the Calogero system.
Submerging it for this purpose in an external auxiliary well
$U^{(A)}(R^{(A)})= A\,\omega^2 [R^{(A)}]^2/8 $ we get a slightly
simplified version of the Calogero Hamiltonian,
 \ben
 \tilde{H}^{(A)}=
 H^{(A)}+
 U^{(A)}
 = \sum_{i=1}^{A}\ \left [-
\frac{\p^2}{\p{x_i}^2}
 + \frac{A}{8}\,\omega^2
\,x_i^2  \right ]
 +\sum_{i<j=2}^{A}\ \frac{g}
 {(x_i-x_j)^{2}}\ .
  \een
This leads to the centre-of-mass equation
 \be
 \left [ -\frac{d^2}{d R^2} +
\frac{A}{8}\,\omega^2R^2-\omega\,\sqrt{ \frac{A}{8}}\, F \right ]
\Psi(R) = 0 \label{SE2cms}
 \ee
with the well known solutions \cite{Fluegge}. Each element of its
spectrum
 \ben
F=F_N = 2N+1, \ \ \ \ \ \ N = 0, 1,
\ldots \
 \een
adds a constant to all the internal energies $E$.



In the first nontrivial model $\tilde{H}^{(2)}$ with the mere two
interacting particles let us ignore the centre-of-mass equation
(\ref{SE2cms}) as trivial and skip $g=0$ (harmonic oscillator).
Our Schr\"{o}dinger bound state problem is then described by the
singular ordinary differential equation in the relative coordinate
$X=(x_1-x_2)/\sqrt{2}$,
 \be
 \left [ -\frac{d^2}{d X^2} +
\frac{1}{4}\,\omega^2X^2+\frac{1}{2}\,\frac{g}{X^{2}}-E \right ]
\psi(X) = 0. \label{SE2}
 \ee
The symbol $E$ denotes the energy in the centre-of-mass system.
The singularity resembles the current centrifugal term with a
non-integer real parameter
 \be
\ell =\ell(g) = -\frac{1}{2} + \sqrt{ \frac{1}{4}+\frac{g}{2} } \
.\label{fu}
 \ee
Such a singularity is, mathematically speaking, too strong. One
has to reduce the domain to a half-axis and fix the ordering of
the particles (say, $x_1 > x_2$, i.e.,  $X> 0$; sometimes, this is
interpreted as a choice of the Bolzmann statistics).

In a mathematically more rigorous setting one should even demand
that the repulsion is not weak, $g \geq 3/2$. Otherwise, the
Hamiltonian admits many self-adjoint extensions, each of which may
lead to a different spectrum \cite{Reed,comment}.

In the Calogero's paper \cite{Calogero} this problem has been
addressed and resolved in a pragmatic spirit. Even when the
irregular solution $\psi^{(-)}(X) \sim |X|^{-\ell}, \ |X| \ll 1$
becomes normalizable (which certainly happens for all $\ell < 1/2$
and/or $g < 3/2$) we eliminate it as ``physically inacceptable"
via an {\em ad hoc} condition. The details of this argument may be
found in the recent comment \cite{comment} where the eligible {\em
ad hoc} conditions were listed as depending on the range of $g$,
 \be
 \begin{array}{ll}
 \lim_{X \to 0} \,[X^{-1/2} \psi^{(Hermitian)}(X)] = 0,
  \ \ \ \ \ \ &
 g \in (-1/2,0),\\
 \lim_{X \to 0} \, \psi^{(Hermitian)}(X) = 0, \ \ \ \ \ \ &
 g \in ( 0, 3/2 ),\\
 \lim_{X \to 0} \psi^{(Hermitian)}(X) = 0, \ \ \ \ \ \ &
 g \in [3/2,\infty).
 \ea
 \label{re}
  \ee
The constraint $g/2 =\ell(\ell+1)> -1/4$ in the first line is
unavoidable and protects the system against a collapse
\cite{Landau}, while only the third line is fully equivalent to
the conventional requirement of normalizability \cite{regul}.

In what follows we shall use the same philosophy. The validity of
the appropriate conditions of the type (\ref{re}) will be
postulated as a conventional regularization of the singularities.
In the spirit of ref. \cite{Calogero} this will define the
``acceptable" solutions and/or make them unique.

In the well known $A=2$ case this implies the termination of the
confluent hypergeometric series to the Laguerre polynomials,
 \be
\psi^{(Hermitian)}_n(X) \sim X^{\ell + 1}
\exp\left(-\frac{1}{4}\,\omega\,X^2 \right) L_n^{\ell+1/2} \left (
\frac{1}{2}\,\omega
 X^2
 \right ), \ \ \ \ \ \ \ \ n = 0, 1, \ldots\ .
\label{waves}
  \ee
Although the two intervals  $X>0$ and $X<0$ are impenetrably
separated, one can deal with the presence of the singularity by
another {\em physically motivated} postulate
 \be
\psi^{(bosonic)}(-X) = + \psi^{(bosonic)}(X), \ \ \ \
\psi^{(fermionic)}(-X) = -\psi^{(fermionic)}(X)
 \label{stathe}
 \ee
mimicking the Bose or Fermi statistics. Such a freedom is rendered
possible by the full separation of the domains $X>0$ and $X<0$. By
construction, the bosonic and fermionic wave functions vanish at
the matching point $X=0$. In what follows, a different type of the
{\em ad hoc} symmetry will be employed and postulated. The
resulting form of the modified statistics will be closely related
to the so called ${\cal PT}$ symmetry in the single-particle
quantum mechanics~\cite{BBjmp}.


\section{${\cal PT} -$symmetric quantum mechanics}

\subsection{$A=1$ and the complexification}

No spikes exist in the trivial Hamiltonian $ \tilde{H}^{(1)}$. It
is equivalent to the harmonic oscillator but offers still a fairly
nontrivial methodical lesson. In their pioneering letter \cite{BB}
Bender and Boettcher emphasized that the complex shift of
coordinates $R =R(r)= r - i\,\varepsilon, \ r \in
(-\infty,\infty)$ in the harmonic oscillator Schr\"{o}dinger
equation (\ref{SE2cms}) does not change its validity {\em and}
preserves the normalizability of its wave functions $\Psi(R)$. The
spectrum remains unchanged even when we admit a symmetric
$r-$dependence in~$\varepsilon=\varepsilon(r^2)$.

Empirically, the similar coexistence of the real spectrum with the
non-Hermitian Hamiltonian has been detected for the various other
single-particle potentials \cite{Caliceti}. The phenomenon has
attracted a lot of attention in the literature \cite{Bessis}. Its
appealing interpretation has been conjectured in terms of the
symbols ${\cal P}$ (which denotes parity, ${\cal P} R {\cal P} =
-R$) and ${\cal T}$ (this is ``time reversal" or complex
conjugation, ${\cal T} i{\cal T}  = -i$). For many Hamiltonians
which commute with the product ${\cal PT} $, people have observed
the reality of the spectra and attributed it {\it expressis
verbis} to the ${\cal PT}$ symmetry \cite{BB,BG,Bessty}.

In the present paper we intend to extend this language to cover
also some systems of more particles.


\subsection{$A =2$ and the regularization }


The complex shift of coordinates did not change the solutions of
$A=1$ equation (\ref{SE2cms}) in the asymptotic region. The same
is true for wave functions of the two Calogero particles. In the
$A=2$ equation (\ref{SE2}) the formula
 \ben
\frac{1}{X^{2}} = \frac{1}{x^2+\varepsilon^2}
-\frac{2\varepsilon^2}{(x^2+\varepsilon^2)^2}
 +\frac{2ix\varepsilon}{(x^2+\varepsilon^2)^2}, \ \ \ \ \ \
 X=x- i\,\varepsilon(x^2)
  \een
indicates how the deformation of the integration contour
regularizes the singularity. An immediate practical compensation
of the loss of the Hermiticity of $ \tilde{H}^{(2)}$ is found in
an improvement of its regularity. This eliminates the rude
constraint (\ref{re}) and opens space for new solutions. At the
same time, we should not get too many of them \cite{Herbst} and so
we demand the ${\cal PT}$ symmetry of the contour,
 \be
 {\cal PT}
 \cdot
 X(x)
 \cdot
 {\cal PT}
= X(-x).
  \label{permu}
 \label{pari}
 \ee
It is possible to search for the even-parity-like solutions of the
complex, regularized radial equation~(\ref{SE2}) in a way proposed
originally by Buslaev and Grecchi  \cite{BG} and dictated by the
angular-momentum interpretation of the parameter $\ell$
\cite{next}. In paper \cite{ptho} we succeeded in complementing
the known ${\cal PT}$-symmetrized solutions (\ref{waves}) of eq.
(\ref{SE2}) by the second hierarchy,
 \be
\psi^{(new)}_n(X) \sim X^{-\ell }
\exp\left(-\frac{1}{4}\,\omega\,X^2 \right) L_n^{-\ell-1/2} \left
( \frac{1}{2}\,\omega
 X^2
 \right ), \ \ \ \ \ \ \ \ n = 0, 1, \ldots\ .
\label{wavesdva}
  \ee
At almost all values of $g$ the pair of solutions (\ref{waves})
and (\ref{wavesdva}) re-connects the subdomains $x>0$ and $x<0$.
The equidistance of the old Hermitian spectrum is manifestly
broken by the new even-like energies. Figure 1 illustrates the
result. It displays both the energies
 \ben
 E_n^{(Hermitian)}=E^{(+)}=\frac{1}{2}\,\omega\,(4n+2\ell+3),
  \ \ \ \ \ \
E^{(new)}_n=E^{(-)}=\frac{1}{2}\,\omega\,(4n-2\ell+1)
 \een
(with $  n = 0, 1, \ldots $) as functions of the coupling $g$ (or
rather $\ell=\ell(g)$) at a fixed value of the spring
constant~$\omega=2\cdot\sqrt{2/A}$.

We may summarize that at $A=2$ the imaginary shifts of the
single-particle coordinates $x_1$ and $x_2$ complexify the Jacobi
coordinates $R$ and $X$. This has comparatively trivial
consequences for the regular centre of mass problem
(\ref{SE2cms}). The change proves much more influential in the
singular radial equation (\ref{SE2}). In the new language we were
able to postulate the two types of  behaviour near $X=0$. {\it
Vice versa}, the current return to the Hermitian constraint
(\ref{re}) acquires now an unconventional understanding of the
reduction of the spectrum, caused by the pragmatic {\em ad hoc}
elimination of all the redundant states~(\ref{wavesdva}).


\section{Generalized statistics}

\subsection{$A=2$ and the idea}

The consistent complexification of our two-body singular equation
(\ref{SE2}) requires an introduction of an upward cut in the
complex plane of $X$.  This reflects the branching role of the
singularity and implies that we have to choose $\varepsilon(0)=
-Im\,X(0)>0$. Still, having the sub-intervals $x> 0$ and $x < 0$
inter-connected, the freedom (\ref{stathe}) of the choice of
statistics seems lost. In fact, it is not. This is to be shown
below.

The analysis is facilitated by the symmetry (\ref{pari}). We can
return to Figure~1 and notice that the levels cross at $g =
2\,k^2-1/2$ for $k = 0, 1, 2,\ldots$. Everywhere off these
exceptional points we can start from the $g=0$ or $\ell =0$ bound
states with the well defined values of parity $=\pm 1$. We propose
to use a smooth continuation in~$g$. This transfers the label $\pm
1$ to almost all the energies. We re-introduce the complexified
bosonic and fermionic symmetry in one of the most natural ways.

In a less intuitive setting we have to imagine that the parity is
not conserved. At $A=2$ we are able to replace this concept by the
equivalent permutation symmetry.  Our knowledge of the explicit
wave functions (\ref{waves}) and (\ref{wavesdva}) enables us to
speak about the bosons and fermions defined by the following rule
 \be
\ba
 \psi^{(bosonic)}(-X)
= (-1)^{-\ell} \psi^{(bosonic)}(X), \ \ \ \ \\
\psi^{(fermionic)}(-X)
= (-1)^{\ell+1}
 \psi^{(fermionic)}(X).
\ea
 \label{statpt}
 \ee
This is a natural generalization of the $\ell = 0$ {\em ad hoc}
conditions (\ref{stathe}) prescribing the behaviour near the
singularity. Mathematically, it enables us to discretize the bound
state spectrum in a way employed also in refs.
\cite{ptho,BG,singular}.

Whenever we choose an integer $\ell$ the role of the parity
becomes partially re-established. The new definition of the
statistics (\ref{statpt}) preserves many features of its Hermitian
predecessor (\ref{stathe}) also in the limit $\lim_{x \to
\infty}\varepsilon(x^2) = 0$. The number of nodal zeros can be
shared by our new bosons and fermions. Their spectra are mutually
shifted. Geometrically they are formed by the opposite branches of
the energy parabolas in Figure~1. In terms of an abbreviation
$\alpha=\ell+1/2$ the transition from fermions to bosons is a mere
change of the sign at this square root, $\alpha =\sqrt{1/4+g/2}
\to -\alpha=-\sqrt{1/4+g/2} $.


\subsection{$A=3$ and a toy model}

The solution of the Calogero three-body problem proves
significantly facilitated after one omits two of its spikes. For
the purely Hermitian toy Hamiltonian
 \ben
 \hat{H}^{(T)}
 = \sum_{i=1}^{3}\ \left [-
\frac{\p^2}{\p{x_i}^2}
 + \frac{3}{8}\,\omega^2
\,x_i^2  \right ]
 +\frac{g}{
 (x_1-x_2)^{2}}\
  \een
this has been noted in the first half of Section 3 in ref.
\cite{Calogero}. After a constant imaginary shift of the variables
$ x_1$, $ x_2$ and $ x_3$ in the three-body Hamiltonian $
\hat{H}^{(T)}$ the centre of mass $R = (x_1+x_2+ x_3)/\sqrt{3}$
becomes complex. Its elimination is still trivial. The other two
Jacobi coordinates
 \ben
 X = \frac{x_1-x_2}{\sqrt{2}}, \ \ \ \ \
Y = \frac{x_1+x_2-2\,x_3}{\sqrt{6}}\
  \een
enter the partial differential Schr\"{o}dinger equation
 \ben
\left [ -\frac{\p^2}{\p X^2} -\frac{\p^2}{\p Y^2}
+\frac{3}{8}\,\omega^2(X^2+Y^2) +\frac{1}{2}g_3X^{-2} -E \right ]
\,\Phi(X,Y)=0.
 \een
At $A=3$ it replaces the $A=2$ radial equation (\ref{SE2}). We
employ its separability in the hyperspherical coordinates $\rho $
and $\phi $. Under the guidance of the Calogeros's paper
\cite{Calogero} this gives $X = \rho\,\sin \phi$ and $Y =
\rho\,\cos \phi$ while $\Phi(X,Y)= \psi(\rho) f(\phi)$. The minor
relevance of the centre-of-mass motion extends naturally to the
$\rho-$dependence. With $\rho \in (0,\infty)$ it is described by
the wave functions
 \ben
\psi_{n,k}(\rho)= \rho^{ \beta(k)} \exp \left (
-\sqrt{\frac{3}{32}}\omega\,\rho^2 \right ) \ L_n^{ \beta(k)}
\left ( \sqrt{\frac{3}{8}}\omega\,\rho^2 \right ), \ \ \ \ \ \ \ \
n, k=0,1,\ldots\
 \een
with the energies
 \ben
E=E_{n,k} =\sqrt{\frac{3}{2}}\,\omega\,[2n+1+ \beta(k)], \ \ \ \ \
\ \ \ n, k=0,1,\ldots\ .
  \een
Both depend on $g$ in a way mediated by the new quantity $
\beta(k)>0$. It is defined as the square root of the eigenvalue of
the ``innermost" hyperangular equation
 \be
\left ( -\frac{d^2}{d \phi^2} + \frac{g}{2\,\sin^2\phi} \right )
\chi_k(\phi) = \beta^2(k)\,\chi_k(\phi).
 \label{coco}
 \ee
Our key idea is to complexify solely the third coordinate $\phi$.
We shall use the recipe
 \ben
\phi = \xi - i\,\varepsilon(\xi), \ \ \ \  \xi \in (-\pi,\pi)
 \een
inspired by its $A=2$ predecessor. The ``Bose" or ``Fermi"
statistics of our toy solutions will be determined by their
regularized behaviour (\ref{statpt}) with $X(x)$ replaced by
$\phi(\xi)$. In effect the complex double well problem
(\ref{coco}) is appropriately constrained in a way consistent with
the permutations of $x_1$ and $x_2$, i.e., with the ${\cal
PT}$-like symmetry
 \ben
 {\cal PT} \cdot
 \phi(
  \xi) \cdot  {\cal PT}=
 \phi(
 -\xi)
 \een
and with the ordinary real parity conservation reflecting the
unconstrained variability of $x_3$,
 \be
 \chi_k(\pi-\phi)=(-1)^k\,\chi_k(\phi)\ .
 \label{lono}
 \ee
This establishes the closest parallels between $A=2$ and $A=3$.


\section{Spectra}

\subsection{Explicit solutions of the toy model}

Differential equation (\ref{coco}) possesses a general solution
which is a hypergeometric series of the Gauss type,
 \ben
\chi^{(\pm)}(\phi)=(\sin\,\phi)^{1/2\pm \alpha} \
_2F_1(u^{(\pm)},v^{(\pm)};1\pm \alpha;\sin^2\phi), \ \ \
\alpha=\frac{1}{2}\sqrt{1+2g}>0
 \een
with $2u^{(\pm)}=1/2- \beta \pm \alpha$ and $2v^{(\pm)}=1/2+ \beta
\pm \alpha$ and with an equivalent alternative form
 \ben
\chi^{(\pm)}(\phi)=\cos \, \phi\,(\sin\,\phi)^{1/2\pm \alpha} \
_2F_1({u}^{(\pm)}+1/2,{v}^{(\pm)}+1/2;1\pm \alpha;\sin^2\phi).
 \een
Only the regular, $^{(+)}-$superscripted states were acceptable in
ref. \cite{Calogero}. For us, the ``irregular" $\chi^{(-)}(\phi)$
represent bosonic states. These functions are smooth and bounded
due to our complex regularization of eq. (\ref{coco}). We can
derive their estimate $ \chi^{(\pm)}(\phi) \sim
\varepsilon^{1/2\pm \alpha}[1+{\cal O}(\varepsilon^2)]$ for the
non-vanishing $\varepsilon \approx \varepsilon(0)> 0$ and in the
closest vicinity of the singularities $\xi = 0$ and $\xi = \pm
\pi$, i.e., for $\xi \in (-\varepsilon, \varepsilon)$, $\xi \in
(-\pi, -\pi+ \varepsilon)$ or $\xi \in (\pi-\varepsilon, \pi)$.
This implies that both the signs can equally well appear at the
parameter~$\alpha$.

On every boundary of convergence $\sin^2 \phi = 1$ of our
power-series solutions we arrive at the same necessity of
termination as before. In the light of the ``real" symmetry
(\ref{lono}) this occurs if an only if our free parameter belongs
to the sequence
 \ben
 \beta(k)=
 \beta^{(\pm)}(k)=
 k \pm \alpha + 1/2,\ \ \ \ \ \ k = 0, 1, \ldots\ . \label{enfi}
  \een
In a way paralleling the Calogero's construction this leads to the
Gegenbauer polynomials in
 \ben
\chi_k^{(fermions/bosons)}(\phi)=(\sin\,\phi)^{1/2\pm \alpha} \
C_k^{1/2\pm \alpha}(\cos \,\phi),\ \ \ \  \ \ \
\alpha=\frac{1}{2}\sqrt{1+2g}>0.
 \een
An analogy of this key formula with its $A=2$ predecessor becomes
clearer in the older notation with $1/2+\alpha=\ell+1$ and
$1/2-\alpha=-\ell$.

The set of the toy eigenvalues $ \beta^2$ decays in the two
subsets. They differ just by the sign attached to the square-root
parameter $\alpha$. The energy spectrum emerges in the compact
form
 \ben
E^{(\pm)}_{n,k}=\sqrt{\frac{3}{8}}\omega\,(4n+2k\pm 2\alpha + 3)
,\ \ \ \  \ \ \ \alpha=\frac{1}{2}\sqrt{1+2g}>0, \ \ \ \ \ n,k=0,
1, \ldots .
 \een
In the convenient scale $\sqrt{\frac{3}{8}}\omega=1$, it is
sampled in Figure~2.


\subsection{The genuine three-body Calogero-type system}

The resemblance between equations (\ref{SE2}) and (\ref{coco}) is
not too surprising since in our toy model the third particle is
only bound by the purely harmonic forces. Let us now recall the
full-fledged Calogero Hamiltonian
 \ben
 \tilde{H}^{(3)}
 = \sum_{i=1}^{3}\ \left [-
\frac{\p^2}{\p{x_i}^2}
 + \frac{3}{8}\,\omega^2
\,x_i^2  \right ]
 +\frac{g}{ (x_1-x_2)^{2}}
 +\frac{g}{ (x_2-x_3)^{2}}
 +\frac{g}{ (x_3-x_1)^{2}}
  \een
and recollect that the related partial differential
Schr\"{o}dinger equation
 \ben
\left [ -\frac{\p^2}{\p X^2} -\frac{\p^2}{\p Y^2}
+\frac{3}{8}\,\omega^2(X^2+Y^2) +\frac{1}{2}g\,X^{-2} \right .
 \een
 \ben
\left . +\frac{1}{2}g\,(\sqrt{3}Y-X)^{-2}
+\frac{1}{2}g\,(\sqrt{3}Y+X)^{-2}-E \right ] \,\Phi(X,Y)=0
 \een
degenerates to the ordinary differential ``innermost" equation
 \ben
M \,f_k(\phi)=
 \beta^2(k)M \,f_k(\phi)
 \een
containing the hyperspherical momentum operator
 \ben M= -\frac{\p^2}{\p \phi^2}
+\frac{1}{2} \left ( \frac{g_{3}}{\sin^2\phi}
+\frac{g_{1}}{\sin^2(\phi+\frac{2}{3}\pi)}
+\frac{g_{2}}{\sin^2(\phi-\frac{2}{3}\pi)} \right ).
 \een
For the equal couplings $g_j=g$ we have the complex differential
equation
 \be
\left ( -\frac{d^2}{d \phi^2} + \frac{9\,g}{2\,\sin^23\,\phi}
\right ) \chi_k(\phi) = \beta^2(k)\,\chi_k(\phi).
 \label{angular}
 \ee
It is a straightforward six-well modification of the above toy
double-well problem with $\xi \in (-\pi,\pi)$. In its analysis let
us start from the Hermitian limit $\varepsilon =0$ with the clear
geometrical meaning of the permutation symmetry in the real $X -
Y$ plane. It is divided into six subdomains separated by the
impenetrable two-body barriers. Thus, the $x_1 \leftrightarrow
x_2$ interchange $ {\cal P}_{(1-2)} $ happens precisely along the
$Y-$axis. Similarly, the other two permutations $x_2
\leftrightarrow x_3$ and $x_3 \leftrightarrow x_1$ take place
along the respective lines $Y=\pm X/\sqrt{3}$ \cite{Calogero}.

As soon as we introduce $\varepsilon \neq 0$ the action of the
complexified permutations has to be subject to the triple rule
 \be
 \ba
 {\cal PT}_{(1-2)} \cdot
 \phi(
  \xi) \cdot  {\cal PT}_{(1-2)}=
 \phi(
 -\xi),\\
 {\cal PT}_{(2-3)} \cdot
 \phi(\xi) \cdot {\cal PT}_{(2-3)} =
 \phi(
  \frac{2}{3}\pi -\xi),\\
 {\cal PT}_{(3-1)} \cdot
 \phi( \xi)  \cdot {\cal PT}_{(3-1)}  =
 \phi(
 -\frac{2}{3}\pi -\xi)
 \ea
 \label{permutri}
 \ee
which guarantees the commutativity of the Hamiltonian with the
vectorial operator $\stackrel{\longrightarrow} {\cal PT}$. In
analogy with equation (\ref{statpt}) we postulate that our
solutions satisfy the six {\em ad hoc} boundary conditions
 \be
\ba
 f^{(bosonic)}(-\phi_j)
 = (-1)^{-\ell}
f^{(bosonic)}(\phi_j), \ \ \ \ \\ f^{(fermionic)}(-\phi_j) =
(-1)^{\ell+1}
 f^{(fermionic)}(\phi_j),\\
 \phi_j = \phi-\frac{j}{3}\pi, \ \ \ \ \ \ j = -2,-1, 0, 1, 2, 3
 .
\ea
 \label{statptri}
 \ee
This extends the ${\cal PT}-$symmetric quantum mechanics of ref.
\cite{BBjmp} to our complexified system of three particles.

\subsection{Solutions}

The parameter $\ell=\ell(g)$ is the same function (\ref{fu}) of
$g$ as above. Incidentally, it vanishes in the harmonic-oscillator
limit $g \to 0$. Explicit solutions of our Calogero-inspired model
can be constructed with an ample use of its toy predecessor. In
particular, the toy hypergeometric solutions have only to be
subject to the modified boundary conditions. This  parallels the
Hermitian situation.

Firstly, due to the overall symmetry (\ref{statptri}) it is still
natural to match the wave functions in the middle of the separate
wells. Formally this relies on the fact that all of the infinite
hypergeometric series reach their limits (i.e., radii) of
convergence precisely at/along the three real-middle-of-the-well
lines $Y=0$ (= $X-$axis) and $Y=\pm \sqrt{3}X$. For this reason
they have to degenerate to the Gegenbauer polynomials as before.

Secondly, the exact values of the energies of the triple-spike
complex Calogero model follow immediately from a mere comparison
of equations (\ref{coco}) and (\ref{angular}). The appropriate
re-scaling of the ``intermediate" eigenvalue parameter $\beta$
gives the resulting final form of the spectrum
 \ben
E^{(\pm)}_{n,k}=\sqrt{\frac{3}{8}}\omega\,(4n+6k\pm 6\alpha + 5)
,\ \ \ \  \ \ \ \alpha=\frac{1}{2}\sqrt{1+2g}>0, \ \ \ \ \ n,k=0,
1, \ldots\ .
 \een
It is illustrated in Figure 3. The pertaining complex angular wave
functions
 \ben
 f^{(fermionic)}_k(\phi)=\chi^{(fermions)}_k(3\phi), \ \ \ \ \ \
 f^{(bosonic)}_k(\phi)=\chi^{(bosons)}_k(3\phi),
 \een
and their oscillator-like radial counterparts
 \ben
\psi^{(\pm)}_{n,k}(\rho)= \rho^{3k\pm 3\alpha+3/2} \exp \left (
-\sqrt{\frac{3}{32}}\omega\,\rho^2 \right ) \ L_n^{
3k\pm 3\alpha+3/2} \left ( \sqrt{\frac{3}{8}}\omega\,\rho^2 \right )
 \een
are easily determined using the elementary insertions in the
obvious manner resembling the classical Calogero's analysis.


\section{Summary}

In a way guided by the available one-body experience we
complexified the two- and three-body Calogero model, having
consequently employed the advantage of its separability. This
enabled us to suppress the majority of complications (e.g., an
interplay between several complex variables) which would
necessarily arise in any more complicated many-body model.

Our conclusion of upmost importance concerns the new role of the
operator ${\cal PT}$. We were guided by its $A=2$ action upon $X
\sim x_1-x_2$ as given by equation (\ref{permu}). This equation
was interpreted as a complexification of the usual permutation of
particles $x_1 \leftrightarrow x_2$. Hence, we also assigned a new
meaning to the whole concept of the ${\cal PT}$ symmetry for more
particles. Its re-examination in the three-body context enabled us
to find its appropriate permutation-symmetry generalization
(\ref{permutri}).

During our application of the complex permutation symmetry to the
Calogero model we arrived at several satisfactory analogies
between its separate $A=1$, $A=2$, and $A=3$ special cases.
Firstly, as long as we studied the complex coordinates only on the
``innermost" level, we were able to leave all the previous stages
(with any complex Jacobi coordinates) open to virtually arbitrary
analytic continuation.

Secondly, it was comparatively easy to specify the correct
physical states in the limit of the standard Calogero system. This
backward relationship to Hermitian predecessors is quite
instructive. Among our bosonic as well as fermionic complex
solutions some of them ``survive" in the limit $\varepsilon \to
0$. Still, as long as the domains themselves become halved, the
full set of solutions is overcomplete. For this reason the ${\cal
PT}$ symmetric bosons were declared to be redundant. They must be
eliminated by brute force. Usually, for physical reasons, their
function in statistics is successfully mimicked by their
non-analytic Calogero substitutes~(\ref{stathe}).

We may summarize that the reality of spectra of our new,
non-Hermitian and complexified Calogero model can be again
tentatively interpreted as stemming from its multiple permutation
$\stackrel{\longrightarrow} {\cal PT}$ symmetry. We can
conjecture, therefore, that this type of a complex permutation
could presumably play the role of a certain ``weakened
Hermiticity" in the many-body physics. This generalizes the parity
$\times$ time-reversal symmetry which is already comparatively
well understood within the ${\cal PT}$ symmetric quantum mechanics
of isolated particles \cite{BBjmp}.

\section*{Acknowledgment}

Our interest in a complexification of few-body systems has been
inspired by our long lasting communication with Alexander
Turbiner.  In particular, his enthusiasm and deep interest in the
solvable Calogero-type models proved contagious, and our numerous
related discussions were extremely illuminating. All this is most
gratefully appreciated. Our further acknowledgements belong to ICN
(UNAM, Mexico) where a part of the work was done, and to the Czech
GA AS (contracts No. A 1048004 and A 1048801).

\newpage

\section*{Figure captions}

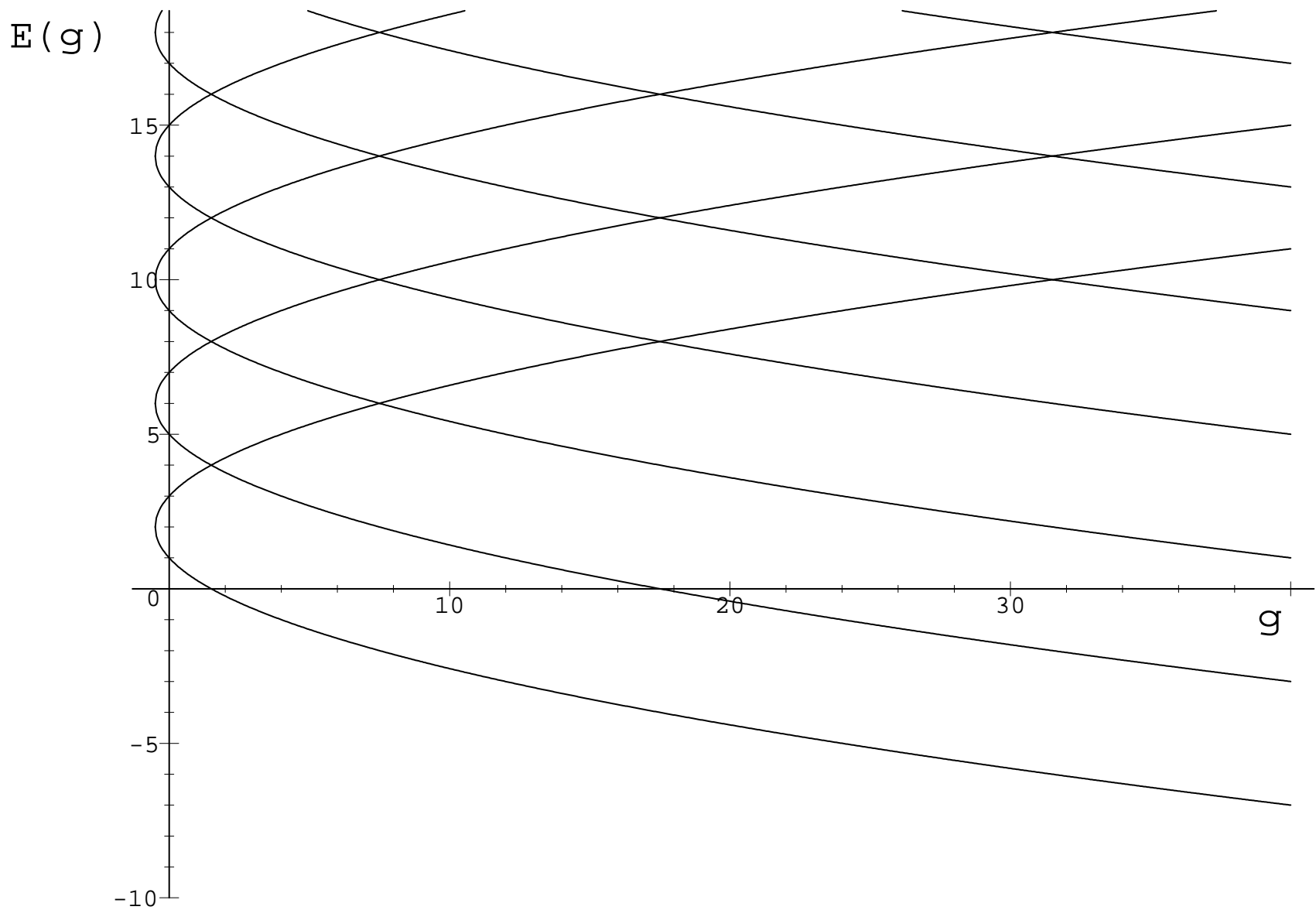
Figure 1. ${\cal PT}$ symmetric energies for $A = 2$

\vspace{5mm}

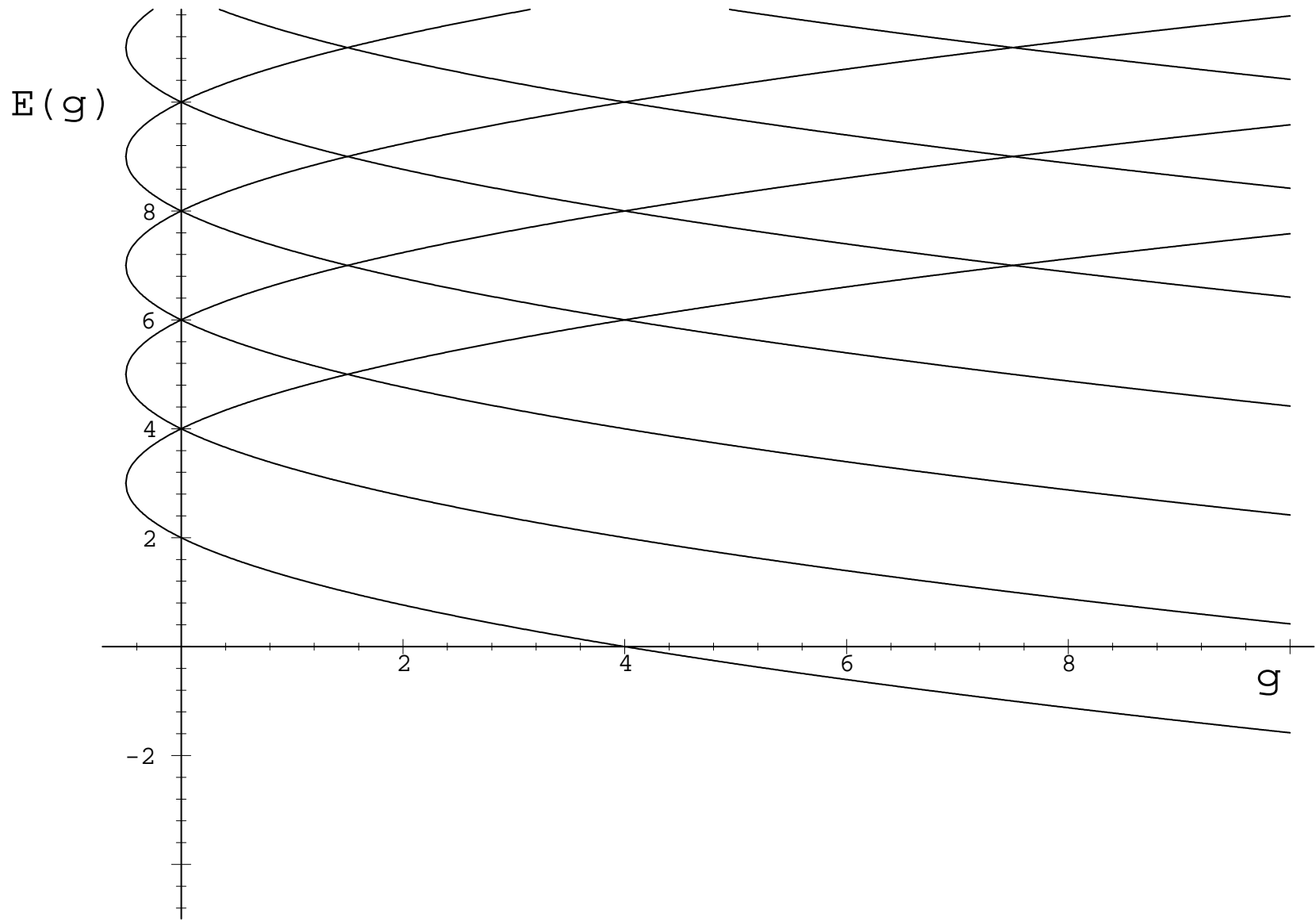
\noindent Figure 2. Toy spectrum

\vspace{5mm}

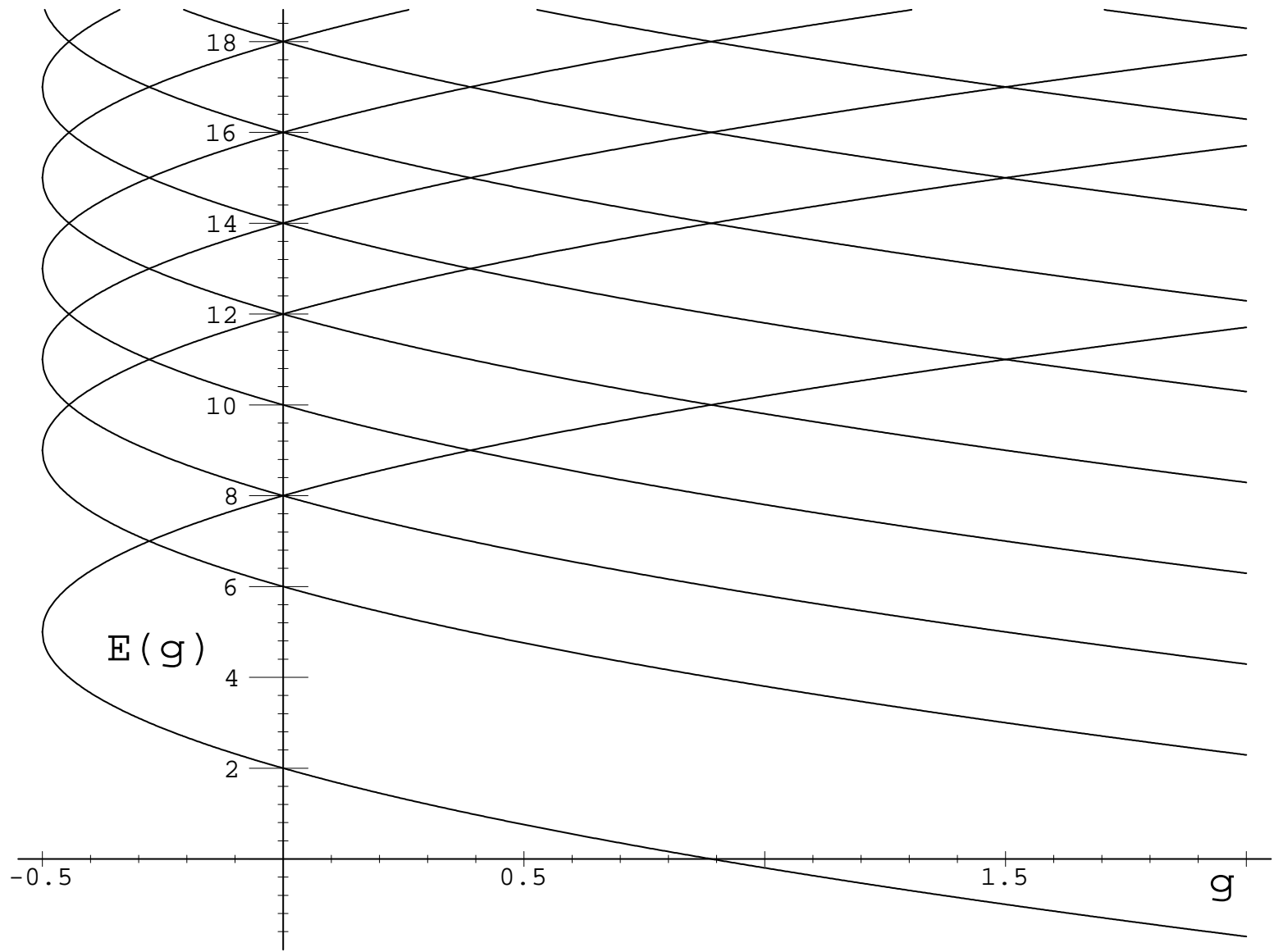
\noindent Figure 3.  ${\cal PT}$ symmetric energies for $A = 3$

\end{document}